# Microwave magnetoelectric fields: An analytical study of topological characteristics

R. Joffe [1,2], R. Shavit [1], and E. O. Kamenetskii [1],

[1] *Microwave Magnetic Laboratory, Department of Electrical and Computer Engineering, Ben Gurion University of the Negev, Beer Sheva, Israel*

[2] *Department of Electrical and Electronics Engineering, Shamoon College of Engineering, Beer Sheva, Israel*

December 10, 2015

**Abstract**
The near fields originated from a small quasi-two-dimensional ferrite disk with magnetic-dipolar-mode (MDM) oscillations are the fields with broken dual (electric-magnetic) symmetry. Numerical studies show that such fields – called the magnetoelectric (ME) fields – are distinguished by the power-flow vortices and helicity parameters [E. O. Kamenetskii, R. Joffe, and R. Shavit, Phys. Rev. E **87**, 023201 (2013)]. These numerical studies can well explain recent experimental results with MDM ferrite disks. In the present paper, we obtain analytically topological characteristics of the ME-field modes. For this purpose, we use a method of successive approximations. In the second approximation we take into account the influence of the edge regions of an open ferrite disk, which are excluded in the first-approximation solving of the magnetostatic (MS) spectral problem. Based on the analytical method, we obtain a "pure" structure of the electric and magnetic fields outside the MDM ferrite disk. The analytical studies can display some fundamental features that are non-observable in the numerical results. While in numerical investigations, one cannot separate the ME fields from the external electromagnetic (EM) radiation, the present theoretical analysis allows clearly distinguish the eigen topological structure of the ME fields. Importantly, this ME-field structure gives evidence for certain phenomena that can be related to the Tellegen and bianisotropic coupling effects. We discuss the question whether the MDM ferrite disk can exhibit properties of the cross magnetoelectric polarizabilities.

## I. INTRODUCTION

The electric displacement current in Maxwell equations allows prediction of wave propagation of electromagnetic fields. In a small ferrite sample (with sizes much less than the free-space electromagnetic wavelength), at ferromagnetic-resonance frequencies, one has negligibly small variation of electric energy. In this case, a dynamical process in a sample is described by three differential equations, without the electric displacement current, [1 – 4]

$$\nabla \cdot \vec{B} = 0, \tag{1}$$

$$\vec{\nabla} \times \vec{E} = -\frac{\partial \vec{B}}{\partial t}, \tag{2}$$

$$\vec{\nabla} \times \vec{H} = 0. \tag{3}$$

A formal use of a set of differential Eqs. (1) – (3), does not allow consideration of any retardation effects. However, without Eq. (2), based on Eqs. (1), (3) and the constitutive relation

$$\vec{B} = \vec{\mu} \cdot \vec{H}, \tag{4}$$

where $\vec{\mu}$ is the permeability tensor, one can obtain solutions for propagating waves inside a ferrite sample. Taking into account the temporal-dispersion properties of a ferrite material at a ferromagnetic resonance, one obtains the Walker equation for magnetostatic-potential (MS-potential) wave function $\psi$ (introduced by a relation $\vec{H} = -\vec{\nabla}\psi$) [5]. The oscillations in small ferrite particles, described by the Walker equation, are called magnetostatic-wave (MS-wave) or magnetic-dipolar-mode (MDM) oscillations [1 – 5].

Interaction of small ferrite particles with electromagnetic radiation is not a trivial problem. MDM oscillations in ferrite spheres excited by external microwave fields were experimentally observed, for the first time, by White and Solt in 1956 [6]. Afterwards, experiments with disk-form ferrite specimens revealed unique spectra of oscillations. While in a case of a ferrite sphere one observed only a few wide absorption peaks of MDM oscillations [6], for a ferrite disk there was a multiresonance (atomic-like) spectrum with very sharp resonance peaks [7 – 9]. Analytically, it was shown [10 – 12] that, contrary to spherical geometry of a ferrite particle analyzed in Ref. [5], the Walker equation (together with the homogeneous boundary conditions for function $\psi$ and its derivatives) for quasi-2D geometry of a ferrite disk gives the Hilbert-space energy-state selection rules for MDM spectra. There are so called $G$-mode spectral solutions [10 – 15].

When we aim to obtain the MDM spectral solutions taking into account also the electric fields in a ferrite disk [to obtain the solutions taking into account Eq. (2)], we have to consider the boundary conditions for a magnetic flux density, $\vec{B} = -\vec{\mu} \cdot \vec{\nabla}\psi$. Analytically, it was shown that in this case (because of specific boundary conditions for a magnetic flux density on a lateral surface of a ferrite disk), one has the helical-mode resonances and the spectral solutions are described by double-valued functions [12, 13]. There are so called $L$-mode spectral solutions. For $L$ modes, the electric field in a vacuum region near a ferrite disk has two parts: $\vec{E} = \vec{E}_c + \vec{E}_p$, where $\vec{E}_c$ is the curl-field component and $\vec{E}_p$ is the potential-field component [15]. While the curl electric field $\vec{E}_c$ in vacuum we define from the Maxwell equation $\vec{\nabla} \times \vec{E}_c = -\mu_0 \frac{\partial \vec{H}}{\partial t}$, the potential electric field $\vec{E}_p$ in vacuum is calculated by integration over the ferrite-disk region, where the sources (magnetic currents $\vec{j}^{(m)} = \frac{\partial \vec{m}}{\partial t}$) are given. Here $\vec{m}$ is dynamical magnetization in a ferrite disk. It was shown that in vacuum near a ferrite disk, the regions with non-zero scalar product $\vec{E}_p \cdot (\vec{\nabla} \times \vec{E}_c)$ can exist [15]. This scalar product is called the field helicity. For time-harmonic fields, the time-averaged helicity factor is expressed as [15]

$$F = \frac{\varepsilon_0}{4} \text{Im}\left[ \vec{E}_p \cdot \left( \vec{\nabla} \times \vec{E}_c \right)^* \right]. \tag{5}$$

MDM oscillations in a quasi-2D ferrite disk are macroscopically coherent quantum states, which experience broken mirror symmetry and also broken time-reversal symmetry [12, 13]. Free-space microwave fields, emerging from magnetization dynamics in quasi-2D ferrite disk, carry orbital angular momentums and are characterized by power-flow vortices and non-zero helicity. Symmetry properties of these fields – called magnetoelectric (ME) fields – are different from symmetry properties of free-space electromagnetic (EM) fields. At the MDM frequency $\omega = \omega_{MDM}$, we have for magnetic induction $\vec{B} = \frac{i}{\omega_{MDM}} \left( \vec{\nabla} \times \vec{E} \right)$. The ME-field helicity density is nonzero only at the resonance frequencies of MDMs and is expressed as

$$F = \frac{\omega_{MDM} \varepsilon_0}{4} \operatorname{Im}\left[ i\vec{E} \cdot \vec{B}^* \right] = \frac{\omega_{MDM} \varepsilon_0}{4} \operatorname{Re}\left[ \vec{E} \cdot \vec{B}^* \right] = \frac{\omega_{MDM}}{4c^2} \operatorname{Re}\left[ \vec{E} \cdot \vec{H}^* \right], \qquad (6)$$

where $c = 1/\sqrt{\varepsilon_0 \mu_0}$. In this equation, both the electric and magnetic fields are potential fields. The parameter defined by Eq. (6) is different from the time-averaged optical (electromagnetic) chirality density, which is obtained for both, the electric and magnetic, curl fields and is expressed as [16 – 19]

$$\kappa = \frac{\omega \varepsilon_0}{2} \operatorname{Im}\left( \vec{E}^* \cdot \vec{B} \right). \qquad (7)$$

The ME-field helicity density $F$ was analyzed in numerical studies [15]. A numerical analysis shows also that distribution of the real power-flow density $\frac{1}{2}\operatorname{Re}\left( \vec{E} \times \vec{H}^* \right)$ of a ME field constitutes the vortex topological structure in vacuum [15, 20]. Also, in Ref. [21] it was shown numerically that together with the real power-flow density, a ME field is characterized by the imaginary $\frac{1}{2}\operatorname{Im}\left( \vec{E} \times \vec{H}^* \right)$ power-flow density.

ME-coupling properties, observed in the near-field structure, are originated from magnetization dynamics of MDMs in a quasi-2D ferrite disk. In general, ME-coupling effects manifest in numerous macroscopic phenomena in solids. Physics underlying these phenomena becomes evident through a symmetry analysis. In isolating crystal materials, in which both spatial inversion and time-reversal symmetries are broken, a magnetic field can induce electric polarization and, conversely, an electric field can induce magnetization [1, 22]. Without requirements of a special kind of a crystal lattice, a ME-coupling term appears in magnetic systems with topological structures of magnetization. In particular, there can be chiral, toroidal, and vortex structures of magnetization [23, 24]. Other examples on a role of magnetization topology in the ME-coupling effects concern orbital magnetization. As it was discussed in Refs. [25, 26], an adequate description of magnetism in magnetic materials should not only include the spin contribution, but also should account for effects originating in the orbital magnetism. It was shown that in the two-dimensional case, orbital magnetization is exhibited due to exceeding of chiral-edge circulations in one direction over chiral-edge circulations in opposite direction [26]. Recently, it was shown that ME coupling can occur also in isotropic dielectrics due to an effect of orbital ME polarizability – topological ME-coupling effect [27 – 29]. In such a case, one has the contribution of orbital currents to the ME coupling. The orbital ME polarizability is due to the pseudoscalar part of the ME coupling and is equivalent to the addition of a term to the electromagnetic Lagrangian – the axion electrodynamics term [30]. That is why the orbital ME response in isotropic dielectrics is referred as the axion orbital ME polarizability [27, 28].

In this paper, we develop a theoretical analysis of the near fields originated from a MDM ferrite disk. The electric and magnetic fields in vacuum are obtained based on the magnetization distributions of MDMs inside a ferrite disk. With use of this model, we study analytically topological characteristics of the ME-field modes. We show that in a vacuum region very near to a ferrite sample there is a good correspondence between the analytical and numerical results for the field structure. For an incident electromagnetic field, the MDM ferrite disk looks as a trap with focusing to a ring, rather than a point. While in numerical investigations, one cannot separate the ME fields from the external EM radiation, the theoretical analysis allows clearly distinguish the eigen topological structure of ME fields. This may concern, in particular, an important question whether the MDM ferrite disk can exhibit properties of the magnetoelectric (cross) polarizabilities. In the paper, we examine the analytically obtained ME-field structure in a view of the Tellegen and bianisotropic coupling effects. Based on the analytical study, we investigate the active and reactive components of the complex power-flow density of the ME fields. The topological structure of ME fields demonstrated by such power flow distributions can explain stability of the eigenstates of MDM oscillations.

## II. THE MODEL AND ANALYTICAL RESULTS FOR THE ME-FIELD STRUCTURES

We use a method of successive approximations. In the first-approximation solving of the MDM spectral problem, we obtain analytically the RF magnetization inside a ferrite disk. The known magnetization and the magnetic current derived from this magnetization are considered as sources for the magnetic and electric fields outside a ferrite disk. Based on the second approximation, we find the magnetic and electric components of the ME field and analyze topology of the ME fields in vacuum. In the theoretical analysis, we use the same disk parameters as in Refs. [14, 15]: the yttrium iron garnet (YIG) disk has a diameter of $2\mathfrak{R} = 3$ mm and the disk thickness is $d = 0.05$ mm; the disk is normally magnetized by a bias magnetic field $H_0 = 4900$ Oe; the saturation magnetization of the ferrite is $4\pi M_s = 1880$ G. We assume that magnetic losses in a ferrite disk are negligibly small.

### A. The first-approximation solutions for MS-potential wave functions

In the first-approximation solution, we ignore the influence of the edge regions outside a quasi-2D ferrite disk [10] (see Fig. 1).

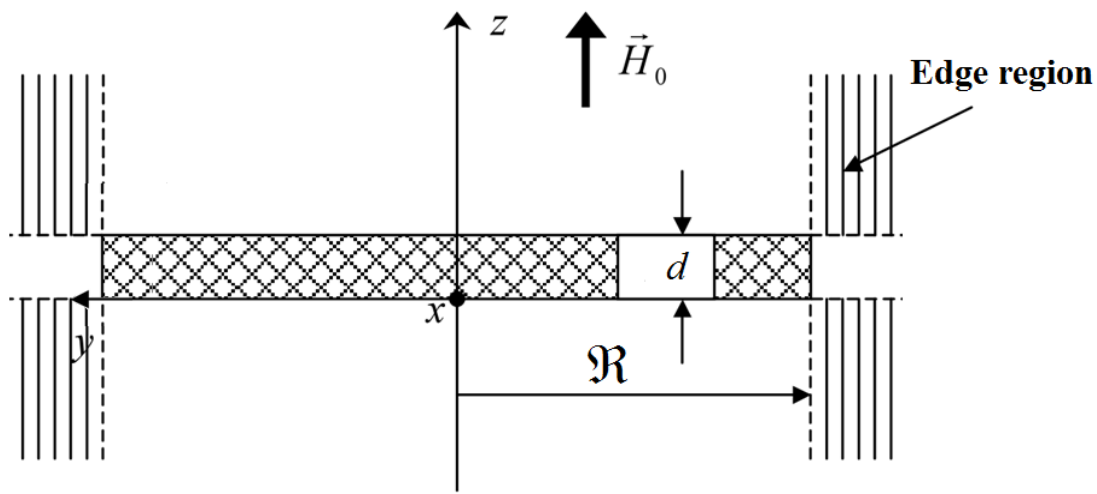

Fig. 1. An open quasi-2D ferrite disk.

Based on this model, we can use separation of variables. Analytically, there are two spectral models for the MDM oscillations. These models give so-called $G$- and $L$-modes. The MS-potential wave function for $L$-mode is written as [10, 12 – 15]

$$\psi_{v,n} = C_{v,n}\xi_{v,n}(z)\tilde{\varphi}_{v,n}(r,\theta), \tag{8}$$

where a dimensionless effective membrane function $\tilde{\varphi}_{v,n}(r,\theta)$ is defined by the Bessel-function orders $v = 1, 2, 3,...$ and the numbers of zeros of the Bessel functions corresponding to different radial variations $n = 1, 2, 3,...$ In Eq. (8), $\xi_{v,n}(z)$ is a dimensionless function of the MS-potential distribution along $z$ axis. $C_{v,n} = QC'_{v,n}$ is an amplitude coefficient where $Q$ is a dimensional unit coefficient of 1A and $C'_{v,n}$ is a normalized dimensionless amplitude. Inside a ferrite disk ($r \leq \Re$, $-d/2 \leq z \leq d/2$) the MS-potential wave function is represented as

$$\psi(r,\theta,z,t) = C_{v,n} J_v\left(\frac{\beta r}{\sqrt{-\mu}}\right)\left(\cos\beta z + \frac{1}{\sqrt{-\mu}}\sin\beta z\right)e^{-iv\theta}e^{i\omega t}, \tag{9}$$

where $\beta$ is a propagation constant for MS waves along z axis and $\mu$ is a diagonal component of the permeability tensor. Solutions in a form of Eq. (9) show the azimuthally-propagating-wave behavior for MS-potential membrane functions.

To define the normalized dimensionless amplitudes $C'_{v,n}$ we will use the Hilbert-space energy-eigenstate $G$-mode spectral solutions. Assuming that the scalar-wave membrane function for the $G$-mode spectral solution is represented as [12 – 15]

$$\tilde{\eta} = \sum_{v,n} C'_{v,n}\tilde{\eta}_{v,n}, \tag{10}$$

we can write

$$\left|C'_{v,n}\right|^2 = \left|\int_S \tilde{\eta}\, \tilde{\eta}^*_{v,n}dS\right|^2. \tag{11}$$

Here integration is made over an entire open-disk region in the $r,\theta$ plane. Normalization of membrane function $\tilde{\eta}$ is expressed as

$$\sum_{v,n}\left|C'_{v,n}\right|^2 = 1. \tag{12}$$

Fig. 2 shows the calculated coefficients $C'_{v,n}$ for the 1st-order Bessel-function ($v = 1$) and the numbers of radial variations ($n = 1, 2, 3,...$).

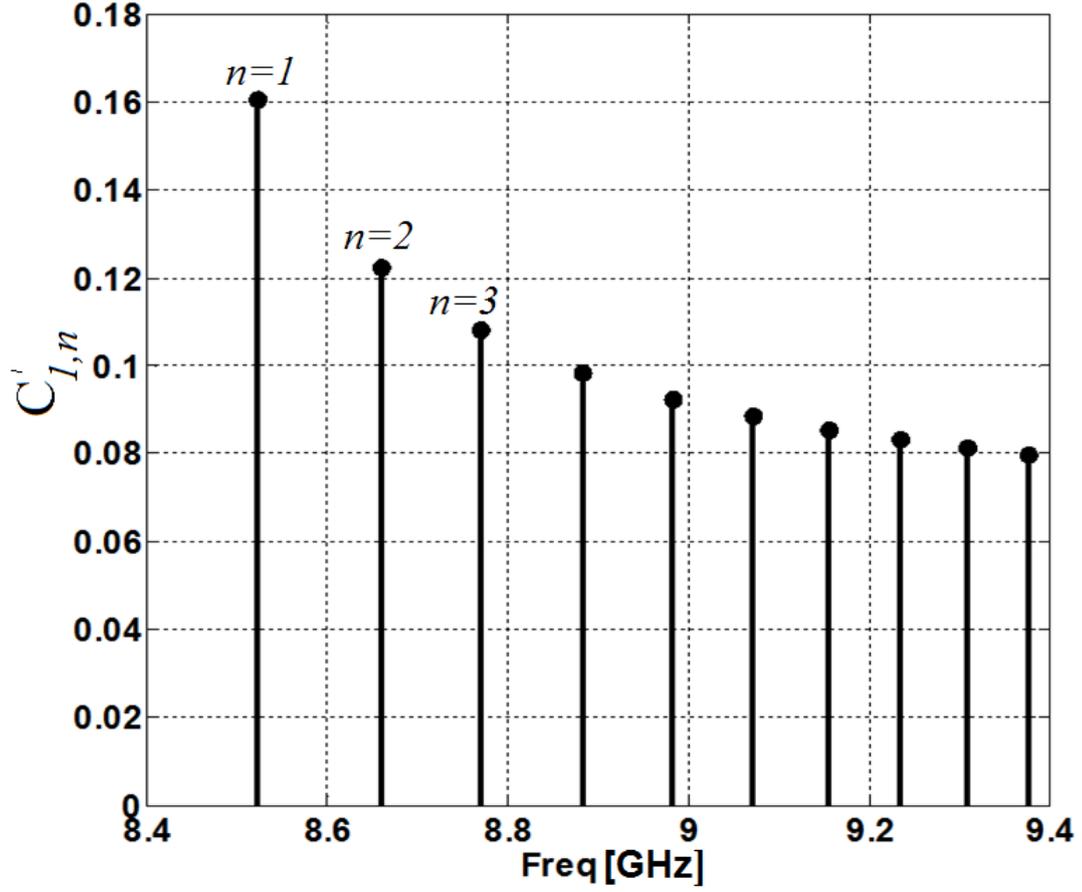

**Fig. 2. Normalized amplitude coefficients for MS-potential wave functions. The MDMs are with the 1$^{st}$-order Bessel-function ($\nu = 1$) and the numbers of radial variations ($n = 1, 2, 3,...$).**

Based on solution (9) for every MDM, we can find the first-approximation solutions for the magnetic field both inside and outside a ferrite disk

$$\vec{H}_{MDM} = -\vec{\nabla}\psi_{MDM}.  \qquad (13)$$

Also, we can find the spectral magnetization distribution inside a ferrite disk:

$$\vec{m}_{MDM} = -\vec{\vec{\chi}} \cdot \vec{\nabla}\psi_{MDM}, \qquad (14)$$

where $\vec{\vec{\chi}}$ is the magnetic susceptibility tensor [4]. For a normally magnetized ferrite disk, we have only the radial and azimuthal components of magnetization. These components are expressed as

$$m_r(r,\theta,z,t) = C_{\nu,n}\left(\cos\beta z + \frac{1}{\sqrt{-\mu}}\sin\beta z\right)\left[\frac{\chi\beta}{\sqrt{-\mu}} J'_\nu\left(\frac{\beta r}{\sqrt{-\mu}}\right) + \frac{\chi_a m}{\rho} J_\nu\left(\frac{\beta r}{\sqrt{-\mu}}\right)\right]e^{-i\nu\theta}e^{i\omega t}, \qquad (15)$$

$$m_\phi(r,\theta,z,t) = -iC_{v,n}\left(\cos\beta z + \frac{1}{\sqrt{-\mu}}\sin\beta z\right)\left[\frac{\chi_a\beta}{\sqrt{-\mu}}J'_v\left(\frac{\beta r}{\sqrt{-\mu}}\right) + \frac{\chi m}{\rho}J_v\left(\frac{\beta r}{\sqrt{-\mu}}\right)\right]e^{-iv\theta}e^{i\omega t}, \quad (16)$$

where $\chi$, $\chi_a$ are diagonal and off-diagonal components of the magnetic susceptibility tensor, respectively [4], and $J'_v$ is a derivative (with respect to the argument) of the Bessel function of order $v$.

Figs. 3 and 4 show MS-potential wave functions for the 1$^{st}$ and 2$^{nd}$ MDMs, respectively.

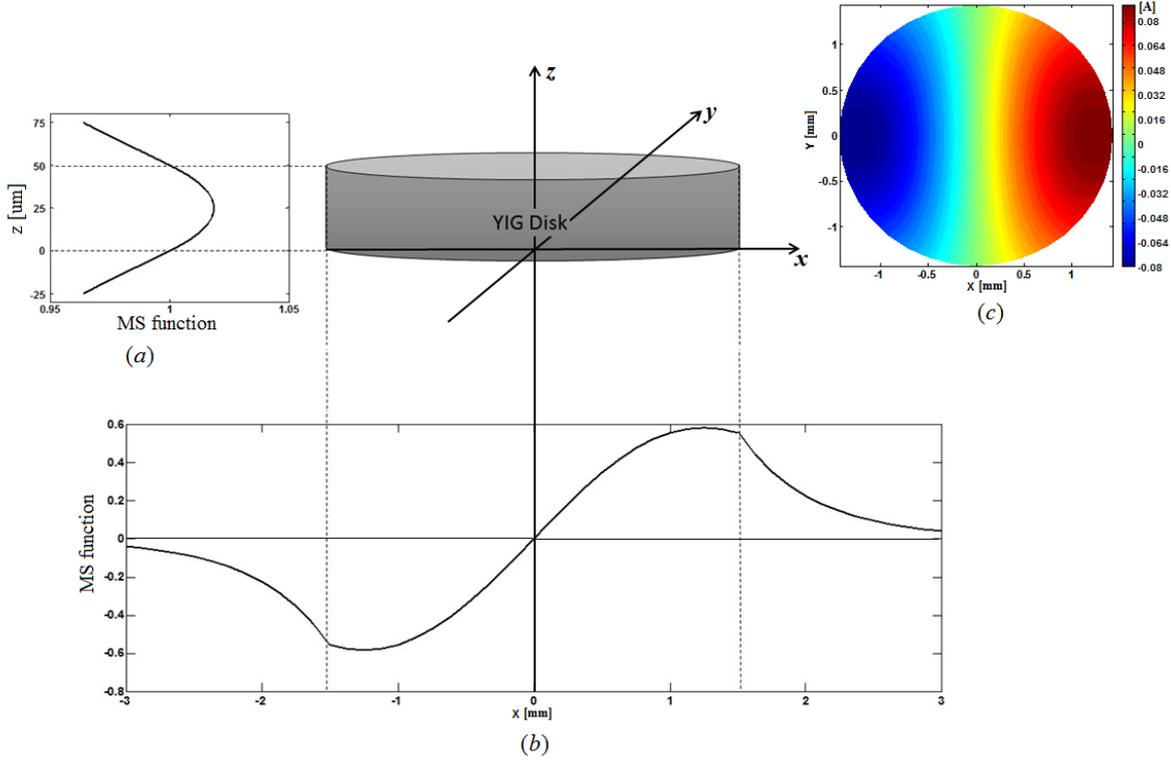

**Fig. 3. MS-potential wave functions (the 1$^{st}$ approximation) for the 1$^{st}$ MDM. The MDM is described with the 1$^{st}$-order Bessel-function ($v=1$) and the radial variation $n=1$. The variation along $z$ axis, $\xi(z)$, corresponds to the main thickness mode. (*a*) The function $\xi(z)$; (*b*) the radial distribution of the MS-potential wave function; (*c*) the picture of intensity of membrane MS-potential wave function $\tilde{\varphi}(r,\theta)$, which rotates in the plane perpendicular to the disk axis *z*.**

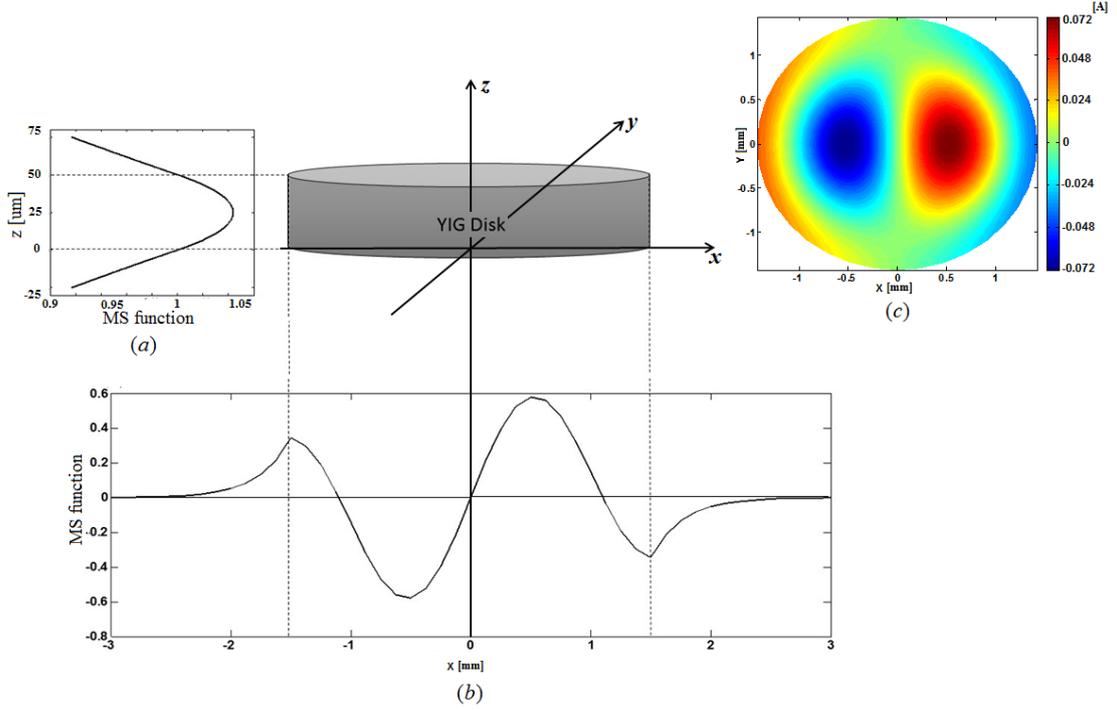

Fig. 4. MS-potential wave functions (the 1$^{st}$ approximation) for the 2$^{nd}$ MDM. The MDM is described with the 1$^{st}$-order Bessel-function ($\nu = 1$) and the radial variation $n = 2$. The variation along z axis, $\xi(z)$, corresponds to the main thickness mode. (*a*) The function $\xi(z)$; (*b*) the radial distribution of the MS-potential wave function; (*c*) the picture of intensity of membrane MS-potential wave function $\tilde{\varphi}(r,\theta)$, which rotates in the plane perpendicular to the disk axis *z*.

These modes are described by the 1$^{st}$-order Bessel-functions ($\nu = 1$). The 1$^{st}$ and 2$^{nd}$ MDMs have the numbers of radial variations $n = 1$ and $n = 2$, respectively. The variation along *z* axis, described by the function $\xi(z)$, corresponds to the main thickness mode. Based on the known MS-potential wave functions, radial and azimuthal components of magnetization were calculated. The in-plane distributions of the magnetization (at the plane $z = d/2$) are shown in Fig. 5 for the 1$^{st}$ and 2$^{nd}$ MDMs.

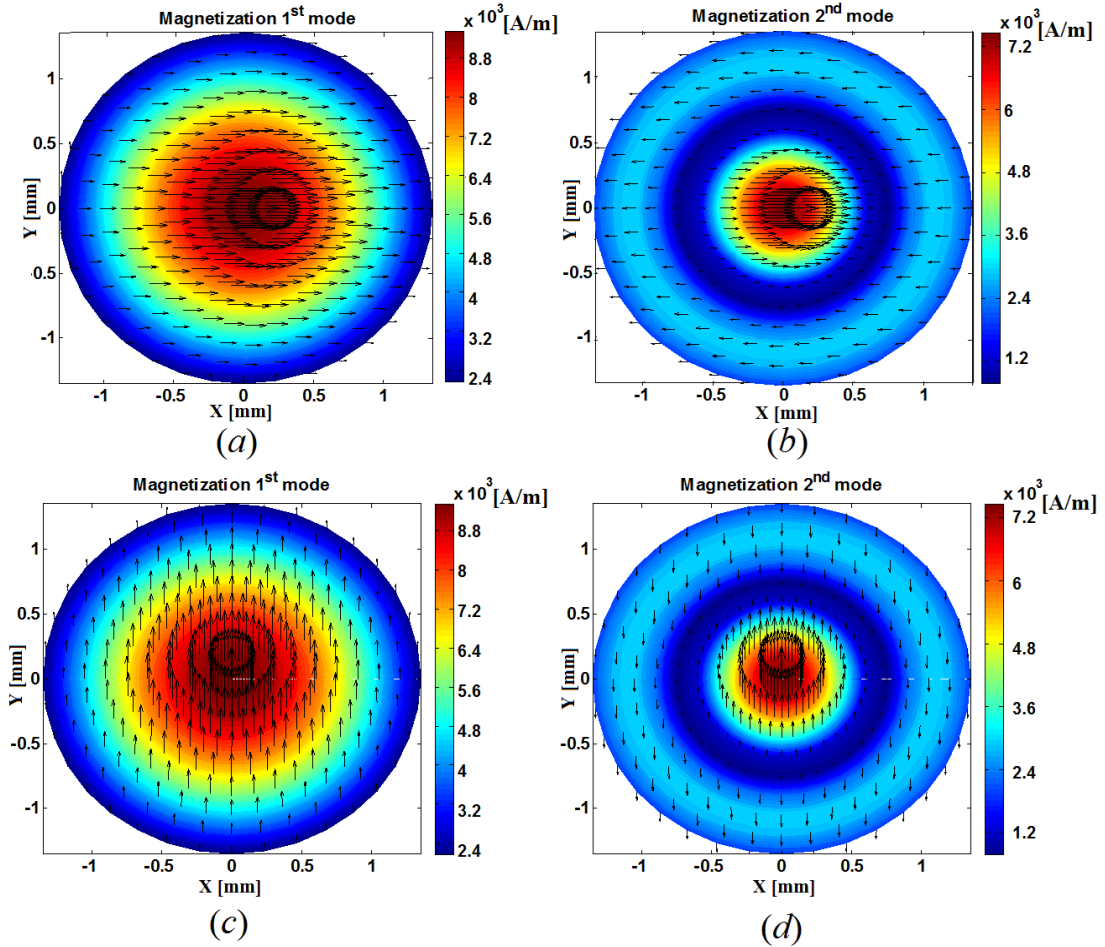

**Fig. 5.** The magnetization distribution in a quasi-2D ferrite disk. (*a*), (*b*) The 1$^{st}$ and 2$^{nd}$ MDMs, respectively, for the phase $\omega t = 0°$; (*c*), (*d*) the 1$^{st}$ and 2$^{nd}$ MDMs, respectively, for the phase $\omega t = 90°$. Every elementary magnetic dipole rotates in the *xy* plane. Because of the azimuthally-propagating-wave behavior for MS-potential membrane functions, the entire pictures of the magnetization rotate as well. Direction of the magnetization rotation, clockwise or counterclockwise, depends on a direction of a bias magnetic field. The arrows are unit vectors.

Every elementary magnetic dipole rotates in the *xy* plane. Because of the azimuthally-propagating-wave behavior for MS-potential membrane functions, the entire pictures of the magnetization rotate as well. Direction of the magnetization rotation, clockwise or counterclockwise, depends on the direction of the bias magnetic field.

### B. The second-approximation solutions

The known, from Eqs. (15), (16), MDM magnetization distributions inside a ferrite disk allow making an analysis by taking into consideration the edge regions, shown in Fig 1. As a result, one can obtain an entire structure of the MS-potential wave function and the fields outside a ferrite disk. There are the second-approximation solutions.

Using Eqs. (1), (13) and taking into account that $\vec{B} = \mu_0(\vec{H} + \vec{m})$, we obtain the following Poisson's equation

$$\nabla^2 \psi = \vec{\nabla} \cdot \vec{m}, \qquad (17)$$

where the right-hand-side term is considered as an effective magnetic charge. If a ferrite disk has a volume *V* and surface *S*, we specify $\vec{m}$ inside *V* as magnetization of a certain MDM and assume that it falls suddenly to zero at the surface *S*. For a given MDM, the solution of Eq. (17) for the MS-potential wave function outside a ferrite disk is [31]

$$\psi(\vec{x}) = -\frac{1}{4\pi}\left(\int \frac{\vec{\nabla}' \cdot \vec{m}(\vec{x}')}{|\vec{x} - \vec{x}'|} dV' - \int \frac{\vec{n}' \cdot \vec{m}(\vec{x}')}{|\vec{x} - \vec{x}'|} dS'\right), \quad (18)$$

where $\vec{n}$ is the outwardly directed normal to surface *S*. The quantity $\vec{n} \cdot \vec{m}$ can be considered as an effective surface magnetic charge density. Because of non-uniform magnetization distributions throughout the volume *V*, both integrals in the right-hand-side of Eq. (18) contribute to the MS-potential solution. Since there are only the radial and azimuthal components of the MDM magnetization, surface *S* in Eq. (18) is a lateral surface of a ferrite disk. Based on Eq. (18), we find the second-approximation solutions for the potential magnetic field outside the ferrite disk:

$$\vec{H}(\vec{x}) = \frac{1}{4\pi}\left(\int \frac{(\vec{\nabla}' \cdot \vec{m}(\vec{x}'))(\vec{x} - \vec{x}')}{|\vec{x} - \vec{x}'|^3} dV' - \int \frac{(\vec{n}' \cdot \vec{m}(\vec{x}'))(\vec{x} - \vec{x}')}{|\vec{x} - \vec{x}'|^3} dS'\right). \quad (19)$$

We can also obtain the second-approximation solutions for the magnetic flux density outside the ferrite disk:

$$\vec{B}(\vec{x}) = \mu_0 \vec{H}(\vec{x}). \quad (20)$$

Fig. 6 shows the normalized magnetic field distributions $\left(\hat{H} = \vec{H}/|\vec{H}|\right)$ outside the ferrite disk (on the *xz* cross-sectional plane) for the 1st and 2nd MDMs.

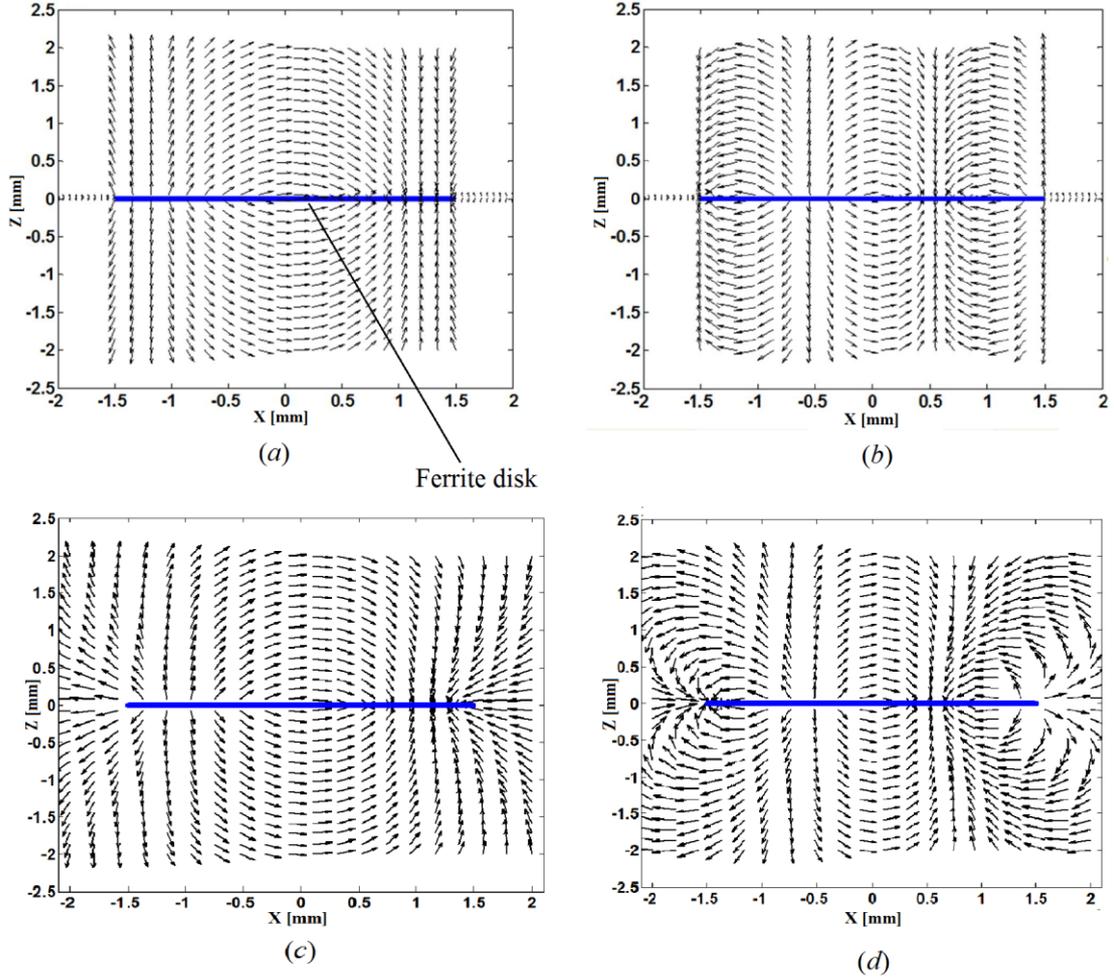

**Fig. 6.** Normalized magnetic field distributions $(\hat{H} = \vec{H}/|\vec{H}|)$ outside a ferrite disk (on the *xz* cross-sectional plane) shown for a certain phase, $\omega t = 0°$. (*a*), (*b*) The 1$^{st}$ and 2$^{nd}$ MDMs, respectively, for the 1$^{st}$ approximation; (*c*), (*d*) the 1$^{st}$ and 2$^{nd}$ MDMs, respectively, for the 2$^{nd}$ approximation. The arrows are unit vectors.

One can compare the magnetic fields calculated based on the 1$^{st}$ approximation [Figs. 6 (*a*) and (*b*)] with the fields calculated based on the 2$^{nd}$ approximation [Figs. 6 (*c*) and (*d*)].

Inside a ferrite disk the Faraday equation (2) is written as

$$\vec{\nabla} \times \vec{E} = -i\omega\mu_0 \vec{H} - \vec{j}^{(m)}, \qquad (21)$$

where we introduced a magnetic current $\vec{j}^{(m)}$, expressed as

$$\vec{j}^{(m)} = i\omega\mu_0 \vec{m}. \qquad (22)$$

Outside a ferrite disk, in vacuum, we consider an electric field as composed by two components: $\vec{E} = \vec{E}_c + \vec{E}_p$. The curl electric field $\vec{E}_c$ is defined as

$$\vec{\nabla} \times \vec{E}_c = -i\omega\mu_0 \vec{H}. \qquad (23)$$

Here $\vec{H}$ is the potential magnetic field in vacuum, found based on Eq. (19). The potential electric field $\vec{E}_p$ in vacuum is originated from a source region: a MDM ferrite disk with a magnetic current $\vec{j}^{(m)}$. Based on an evident duality with the classical-electrodynamics problem of the magnetostatic magnetic field originated from the electric-current source region [31], in Ref. [15] it was shown that the potential electric field $\vec{E}_p$ outside the ferrite disk is defined as

$$\vec{E}_p(\vec{x}) = -\frac{1}{4\pi} \int \frac{\vec{j}^{(m)}(\vec{x}') \times (\vec{x} - \vec{x}')}{|\vec{x} - \vec{x}'|^3} dV'. \qquad (24)$$

For every MDM, the potential electric field $\vec{E}_p$ outside the ferrite disk is obtained based on Eq. (24). A magnetic current $\vec{j}^{(m)}$ is calculated based on Eqs. (15), (16), (22).

Fig. 7 shows the normalized electric field distributions $(\hat{E} = \vec{E}/|\vec{E}|)$ outside the ferrite disk (on the *xz* cross-sectional plane) for the 1st and 2nd MDMs.

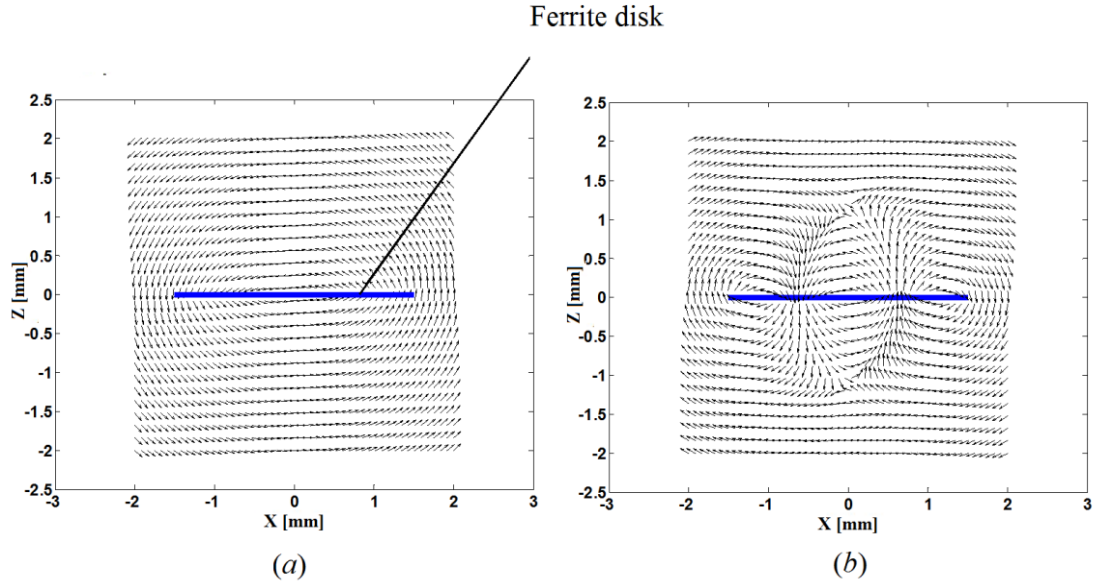

**Fig. 7. Distributions of the normalized potential electric field $(\hat{E} = \vec{E}/|\vec{E}|)$ outside a ferrite disk (on the *xz* cross-sectional plane) shown for a certain phase, $\omega t = 0°$. (*a*) The 1st MDM; (*b*) the 2nd MDM. The arrows are unit vectors.**

In Fig. 8 one can see the calculated distributions of the magnetic and electric fields for the 1st and 2nd MDMs on a vacuum *xy* plane 20 um above the ferrite disk.

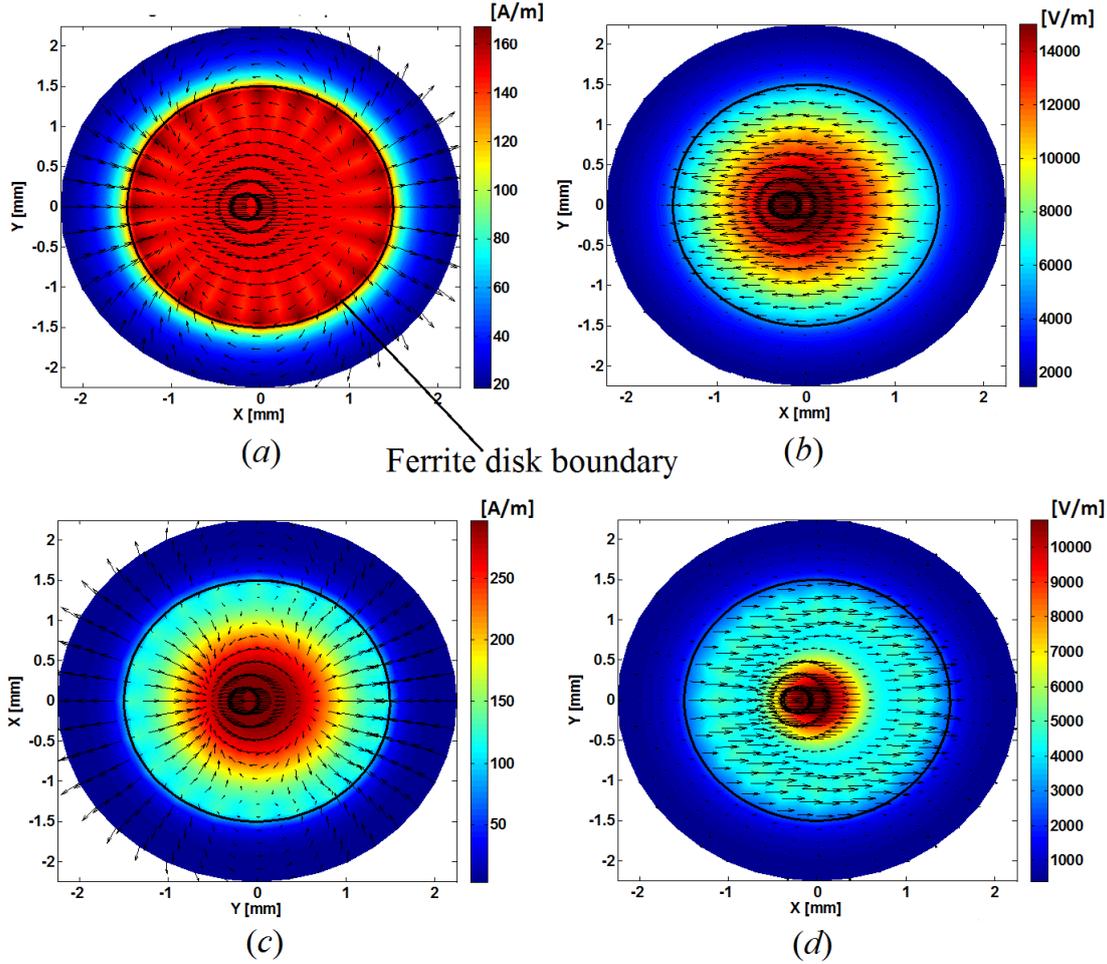

**Fig. 8.** The field distributions on a vacuum *xy* plane 20 um above a ferrite disk shown for a certain phase, $\omega t = 0°$. (*a*), (*b*) The magnetic and electric fields, respectively, for the 1$^{st}$ MDM; (*c*), (d) the magnetic and electric fields, respectively, for the 2$^{nd}$ MDM. The arrows are unit vectors.

In these figures we present only the distributions of the potential fields $\vec{E}_p$ for the 1$^{st}$ and 2$^{nd}$ MDMs. A comparative analysis of the analytically derived distribution of the potential fields $\vec{E}_p$ and numerically obtained distribution of the total field $\vec{E} = \vec{E}_p + \vec{E}_c$, shown below in Figs. 15 and 16, gives evidence for a negligibly small role of the curl electric field $\vec{E}_c$ in the field topology. Theoretically, the potential electric field $\vec{E}_p$ in a vacuum region near a ferrite disk is found by integration over a ferrite-disk volume for a known magnetic current $\vec{j}^{(m)}$ [Eq. (24)]. To obtain the curl electric field $\vec{E}_c$, one has to solve the differential equation (23) taking into account that the magnetic field in a vacuum region is the evanescent-wave field. For geometry of an open ferrite disk [with taking into consideration the edge regions (see Fig. 1)], analytical solutions of Eq. (23) is associated with substantial difficulties. Because of a negligibly small role of the curl electric field $\vec{E}_c$ in the near-field topology, an analysis of such a component of the electric field is beyond the frames of this work.

## C. Topological characteristics of ME fields

Based on the known analytical solutions for the MS-potential wave function and the fields outside the ferrite disk, we can obtain topological characteristics of the ME fields. The distributions of the helicity density and the power-flow density comprise these characteristics. The ME-field helicity density, expressed by Eq. (6), was calculated based on Eqs. (19) and (24) for potential magnetic and electric fields. Fig. 9 shows the calculated ME-field helicity density distributions for the 1$^{st}$ and 2$^{nd}$ MDMs. One can see that a dimension of the ME-field helicity density $F$ is Joule/meter$^4$. For a potential field, this is a gradient of energy density which defines field strength. So, in our case, factor $F$ is a strength of a ME field. For potential electric and magnetic fields, magnitudes and directions of the field strengths are determined by forces acting on a test electric charge or an electric current element. The question what kind of a test element should be introduced to determine the strength of the ME field is open. This question is beyond an analysis in the present paper.

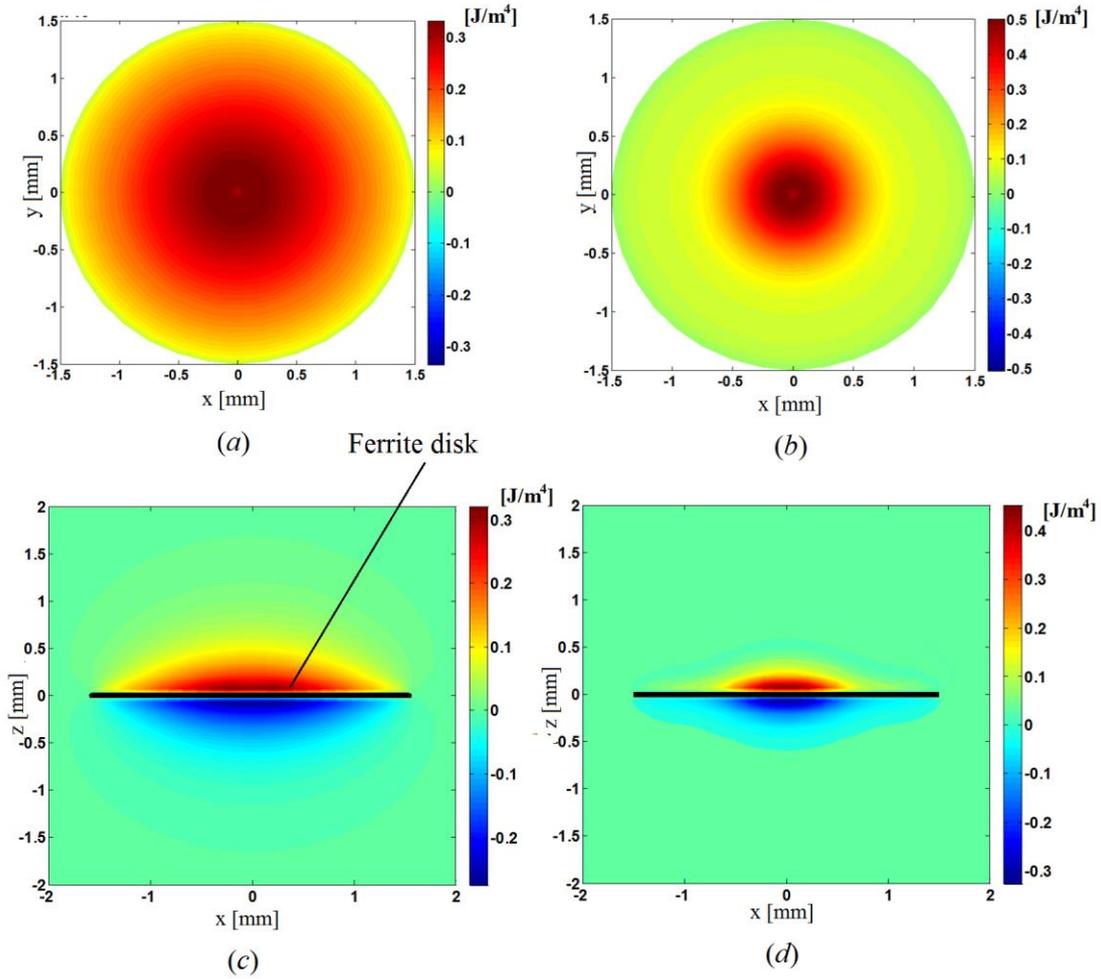

**Fig. 9. The ME-field helicity density analytically calculated based on Eq. (6). (*a*) and (*b*) The distributions on a vacuum *xy* plane 20 um above a ferrite disk for the 1$^{st}$ and 2$^{nd}$ MDMs, respectively; (*c*) and (*d*) the distributions on a cross-sectional *xz* plane for the 1$^{st}$ and 2$^{nd}$ MDMs, respectively.**

In an analysis of topological properties of ME fields, an important question arises what is a distribution of an angle between the potential electric and magnetic fields in the near-field space. While for regular EM-field problems, the vectors of electric and magnetic fields in vacuum are mutually perpendicular in space, in our case, a space angle between the electric and magnetic fields in vacuum is, definitely, not 90°. An angle between the potential electric and magnetic fields we will characterize by the parameter called the normalized ME-field helicity:

$$\cos\alpha_p = \frac{\text{Re}\left(\vec{E}_p \cdot \vec{H}^*\right)}{\left|\vec{E}_p\right|\left|\vec{H}\right|}. \tag{25}$$

Here subscript $p$ for an angle $\alpha$ means "potential". The distributions of $\cos\alpha_p$ on a cross-sectional $xz$ plane are shown in Fig. 10 for the 1st and 2nd MDMs.

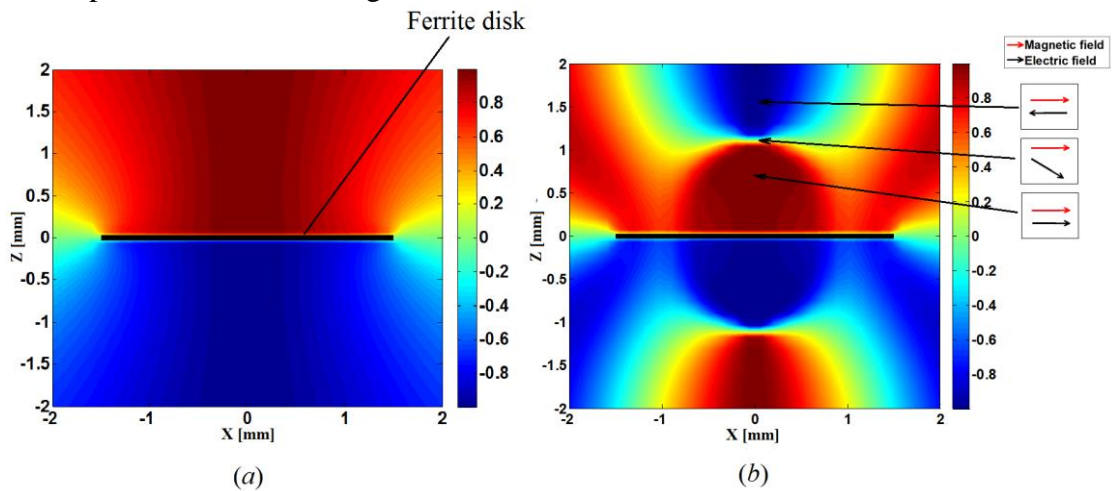

**Fig. 10. The normalized ME-field helicity ($\cos\alpha_p$) distributions on a cross-sectional $xz$ plane. (*a*) For the 1st MDM; (*b*) for the 2nd MDM. The insertions illustrate mutual orientations of the electric and magnetic fields in different points on the $z$ axis above a ferrite disk.**

In fact, this parameter represents mutual orientations of the electric- and magnetic-field vectors shown in Figs. 6 (c, d), 7, and 8.

In an analysis of the complex power-flow density of the ME fields, we will use the following consideration. A simple analysis of the energy balance equation for monochromatic MS waves in a magnetic medium with small losses [12] shows that the real power flow density for a MDM is expressed as

$$\vec{p}_s = \text{Re}\left(i\omega\psi_s^*\vec{B}_s\right) = \frac{i\omega}{2}\left(\psi_s^*\vec{B}_s - \psi_s\vec{B}_s^*\right). \tag{26}$$

Here the mode number $s$ includes both the Bessel-function orders $\nu$ and the numbers of zeros of the Bessel functions corresponding to different radial variations $n$. With use of the first-approximation solution for the MS-potential wave function, expressed by Eq. (9), we can calculate the quantity $\vec{p}_s$. It is easy to show [32] that inside a ferrite disk $(p_s)_r = (p_s)_z = 0$. The only non-zero component of the real power flow density inside a ferrite disk is the azimuth component, which is expressed as [21, 32]:

$$
\begin{aligned}
\left(p_s(r,z)\right)_\theta &= \frac{i\omega}{2} C_s^2 \left(\xi(z)\right)^2 \left[ -\mu \frac{1}{r}\left( \tilde{\varphi}_s^* \frac{\partial \tilde{\varphi}_s}{\partial \theta} - \tilde{\varphi}_s \frac{\partial \tilde{\varphi}_s^*}{\partial \theta} \right) + i\mu_a \left( \tilde{\varphi}_s^* \frac{\partial \tilde{\varphi}_s}{\partial r} + \tilde{\varphi}_s \frac{\partial \tilde{\varphi}_s^*}{\partial r} \right) \right] \\
&= -\omega\, C_s^2 \left(\xi_s(z)\right)^2 \left[ \frac{\mu}{r} \nu \left|\tilde{\varphi}_s(r,\theta)\right|^2 + \frac{1}{2}\mu_a \frac{\partial \left|\tilde{\varphi}_s(r,\theta)\right|^2}{\partial r} \right],
\end{aligned}
\tag{27}
$$

where $\mu$ and $\mu_a$ are, respectively, the diagonal and off-diagonal components of the tensor $\ddot{\mu}$. The quantities $\left(p_s(r,z)\right)_\theta$, circulating around a circle $2\pi r$ (where $r \leq \Re$), are the MDM power-flow-density vortices with cores at the disk center. At a vortex center the amplitude of $(p_s)_\theta$ is equal to zero. In a vacuum region, outside a ferrite disk we have from Eq. (27)

$$
\left(p_s(r,z)\right)_\theta = -\omega\, C_s^2 \frac{1}{r}\nu \left(\xi_s(z)\right)^2 \left|\tilde{\varphi}_s(r,\theta)\right|^2. \tag{28}
$$

It is clear that $\vec{\nabla} \cdot \vec{p}_s = 0$, both inside and outside a ferrite disk.

In Ref. [21], an analysis of the energy relations for MDMs was extended by an introduction also the imaginary power flow density for a MDM. For a mode $s$, this imaginary power flow density is expressed as

$$
\vec{q}_s = \mathrm{Im}\left(i\omega \psi_s^* \vec{B}_s\right) = \frac{\omega}{2}\left(\psi_s^* \vec{B}_s + \psi_s \vec{B}_s^*\right). \tag{29}
$$

Similarly to the above calculation of the imaginary power flow density $\vec{p}_s$, we can calculate the imaginary power flow density $\vec{q}_s$ based on the first-approximation solution (9). In this case, we can show that inside a ferrite that $(q_s)_\theta = 0$, while $(q_s)_r \neq 0$ and $(q_s)_z \neq 0$. For $(q_s)_r$, we obtain:

$$
\begin{aligned}
\left(q_s(z)\right)_r &= -\frac{\omega}{2} C_s^2 \left(\xi(z)\right)^2 \left[ \mu\left( \tilde{\varphi}_s^* \frac{\partial \tilde{\varphi}_s}{\partial r} + \tilde{\varphi}_s \frac{\partial \tilde{\varphi}_s^*}{\partial r} \right) + i\mu_a \frac{1}{r}\left( \tilde{\varphi}_s^* \frac{\partial \tilde{\varphi}_s}{\partial \theta} - \tilde{\varphi}_s \frac{\partial \tilde{\varphi}_s^*}{\partial \theta} \right) \right] \\
&= -\omega C_s^2 \left(\xi(z)\right)^2 \left[ \frac{1}{2}\mu \frac{\partial \left|\tilde{\varphi}_s(r)\right|^2}{\partial r} - \mu_a \frac{\nu}{r}\left|\tilde{\varphi}_s(r)\right|^2 \right].
\end{aligned}
\tag{30}
$$

Outside the ferrite disk we have

$$
\left(q_s(\theta,z)\right)_r = -\frac{\omega}{2} C_s^2 \left(\xi(z)\right)^2 \frac{\partial \left|\tilde{\varphi}_s(r)\right|^2}{\partial r}. \tag{31}
$$

For $(q_s)_z$, we obtain

$$
\left(q_s\right)_z = -\omega C_s^2 \xi_s(z) \frac{\partial \xi_s(z)}{\partial z}\left|\tilde{\varphi}_s(r,\theta)\right|^2. \tag{32}
$$

The above results for the complex power-flow density of the ME fields, obtained based on the first-approximation solutions, can be improved with use of the second-approximation solutions for MS-potential wave functions, in which an entire structure of the fields outside a ferrite disk is taken into account. For the second-approximation solutions, quantities of the real power flow density $\vec{p}_s$ and the imaginary power flow density $\vec{q}_s$ outside a ferrite disk are calculated based on Eqs. (26) and (29) using Eqs. (18) and (20) for the MS-potential wave function and the magnetic flux density, respectively. Figs. 11 and 12 show the ME-field active and reactive power-flow distributions outside a ferrite disk for the 1$^{st}$ and 2$^{nd}$ MDMs calculated based on Eqs. (26) and (29).

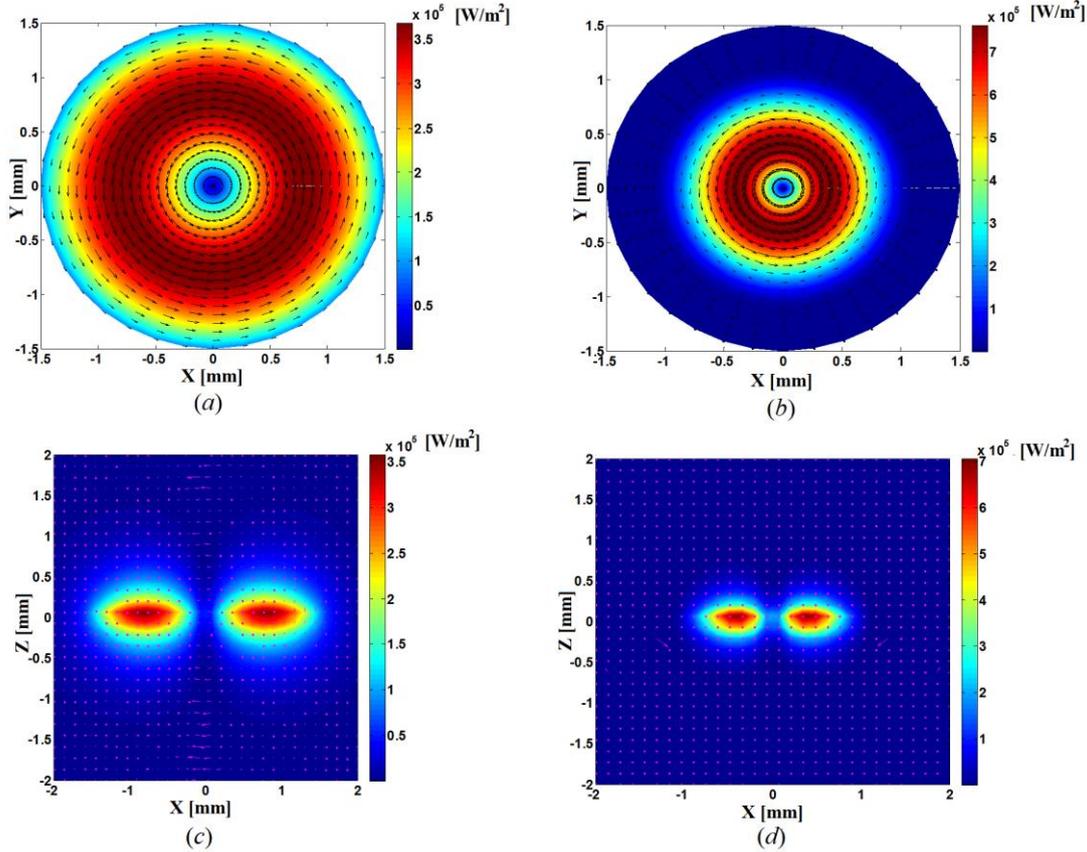

**Fig. 11. The ME-field active-power-flow distribution outside a ferrite disk calculated based on Eq. (26).** (*a*) and (*b*) **The distributions on a vacuum *xy* plane 20 um above a ferrite disk for the 1$^{st}$ and 2$^{nd}$ MDMs, respectively;** (*c*) and (d) **the distributions on a cross-sectional *xz* plane for the 1$^{st}$ and 2$^{nd}$ MDMs, respectively. The arrows are unit vectors.**

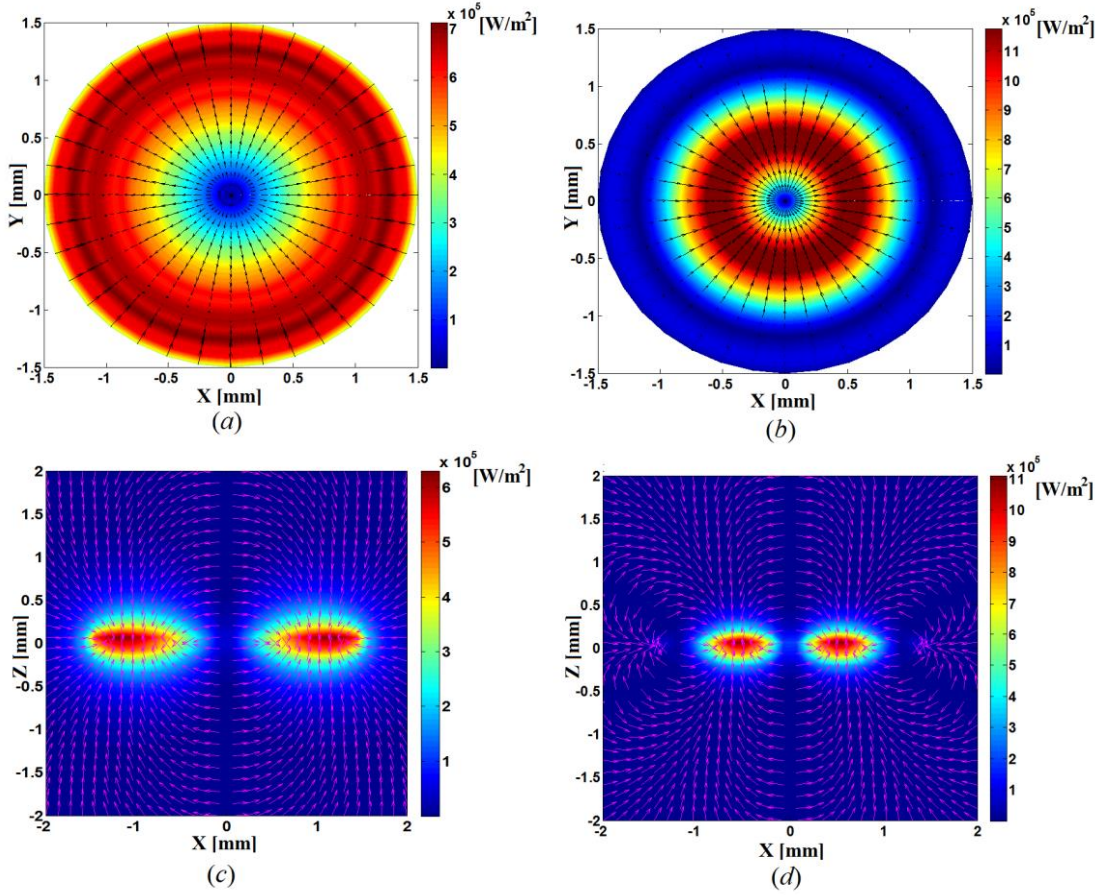

Fig. 12. The ME-field reactive-power-flow distribution outside a ferrite disk calculated based on Eq. (29). (*a*) and (*b*) The distributions on a vacuum *xy* plane 20 um above a ferrite disk for the 1$^{st}$ and 2$^{nd}$ MDMs, respectively; (*c*) and (*d*) the distributions on a cross-sectional *xz* plane for the 1$^{st}$ and 2$^{nd}$ MDMs, respectively. The arrows are unit vectors.

The real and imaginary power flow densities, defined by Eqs. (26) and (29), are related to the time averaged real and imaginary parts of a vector product of the curl electric and potential magnetic fields. A simple manipulation (taking into account that $\vec{\nabla} \cdot \vec{B} = 0$ and $\vec{H} = -\vec{\nabla}\psi$) shows that

$$\vec{\nabla} \cdot (\vec{E}_c \times \vec{H}^*) = \vec{H}^* \cdot \vec{\nabla} \times \vec{E}_c = i\omega \vec{\nabla}\psi^* \cdot \vec{B} = i\omega \vec{\nabla} \cdot \left(\psi^* \vec{B}\right). \tag{33}$$

Following Ref. [33], one can conclude that this equation gives

$$\vec{E}_c \times \vec{H}^* = i\omega\psi^* \vec{B}. \tag{34}$$

For mode *s*, we have

$$\vec{p}_s = \mathrm{Re}\left(i\omega\psi_s^* \vec{B}_s\right) = \mathrm{Re}\left[\left(\vec{E}_c\right)_s \times \left(\vec{H}^*\right)_s\right] \tag{35}$$

and

$$\vec{q}_s = \text{Im}\left(i\omega\psi_s^* \vec{B}_s\right) = \text{Im}\left[\left(\vec{E}_c\right)_s \times \left(\vec{H}^*\right)_s\right]. \tag{36}$$

Importantly, despite the fact that such expressions as $\text{Re}\left(\vec{E}\times\vec{H}^*\right)$ and $\text{Im}\left(\vec{E}\times\vec{H}^*\right)$ looks like the real and imaginary Poynting vectors, the MS-wave power flow densities cannot be basically related to the EM-wave power flow densities. The Poynting vector is obtained for EM radiation which is described by the *two curl operator* Maxwell equations for the electric and magnetic fields [31]. This is not the case described by Eq. (33), where, for the MS waves, we have *potential magnetic* and *curl electric* fields.

While Eqs. (35) and (36), are relevant only for a curl electric field $\vec{E}_c$ in a vacuum region near a ferrite disk, the question about a role of the potential electric field $\vec{E}_p$ in the vector-product quadratic relations arises as well. Formally, we can extend our analysis of the active power flow also to calculations of the quantity $\text{Re}\,\vec{E}_p \times \vec{H}^*$. This quantity, for the 1st and 2nd MDMs, is shown in Fig. 13. Similar to the results shown in Fig. 11, in the picture in Fig. 13 we can see that there are the power-flow-density vortices with cores at the disk center. At a vortex center the amplitude is equal to zero.

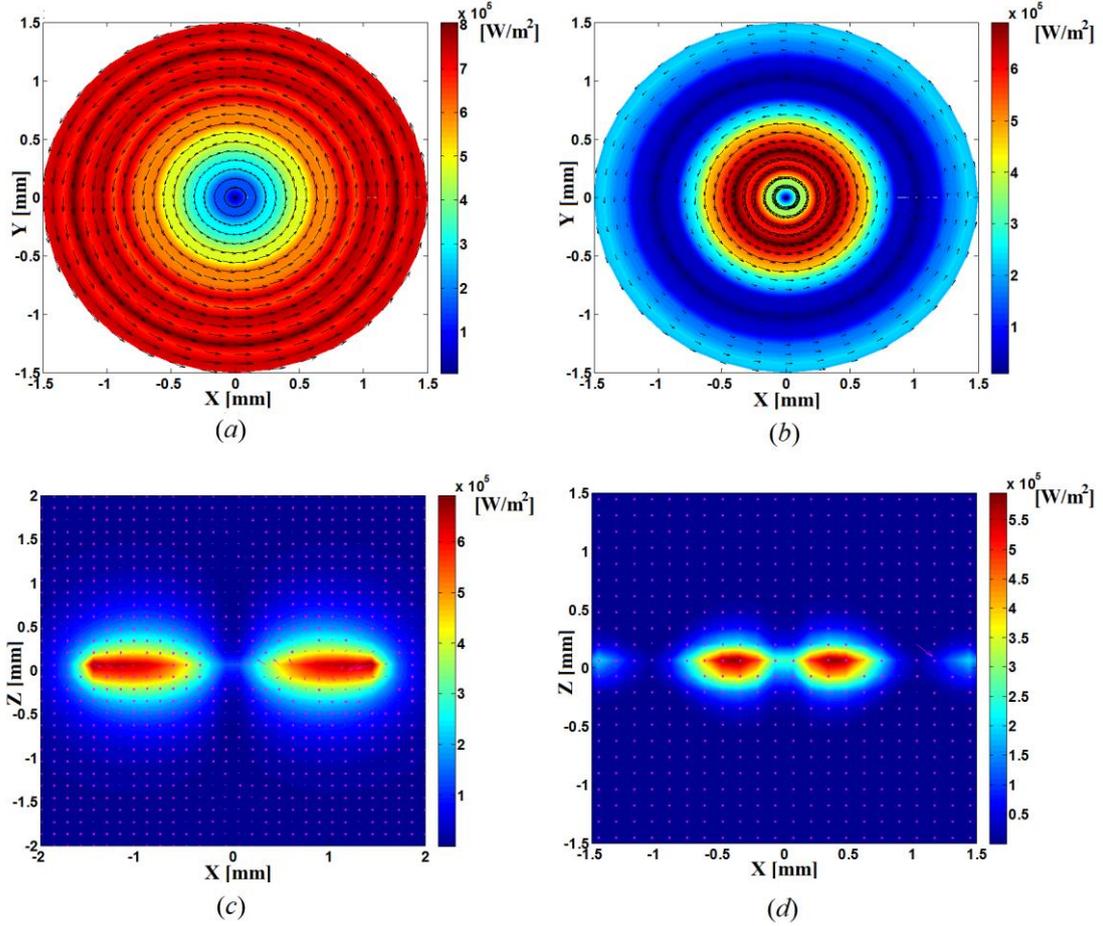

**Fig. 13.** The quantity $\text{Re}\,\vec{E}_p \times \vec{H}^*$ outside a ferrite disk. (*a*) and (*b*) The distributions on a vacuum *xy* plane 20 um above a ferrite disk for the 1st and 2nd MDMs, respectively; (*c*) and (*d*) the distributions on a cross-sectional *xz* plane for the 1st and 2nd MDMs, respectively. The arrows are unit vectors.

In vacuum, near the central region of the MDM ferrite disk, one has in-plane rotating vectors of the potential electric and magnetic fields [15]. It was shown [21] that for these spinning electric- and magnetic-field vectors, a time averaged imaginary part of a vector product of the electric and magnetic fields is related to the reactive power flow density. Based on Eq. (19) and (24), we can find analytically the distribution of the quantity $\operatorname{Im} \vec{E}_p \times \vec{H}^*$ outside the ferrite disk. Fig. 14 shows the calculated quantity of the reactive power flow $\operatorname{Im} \vec{E}_p \times \vec{H}^*$ outside the ferrite disk for the 1st and 2nd MDMs. Contrary to $\operatorname{Re} \vec{E}_p \times \vec{H}^*$, the quantity $\operatorname{Im} \vec{E}_p \times \vec{H}^*$ has a maximal amplitude at the disk center.

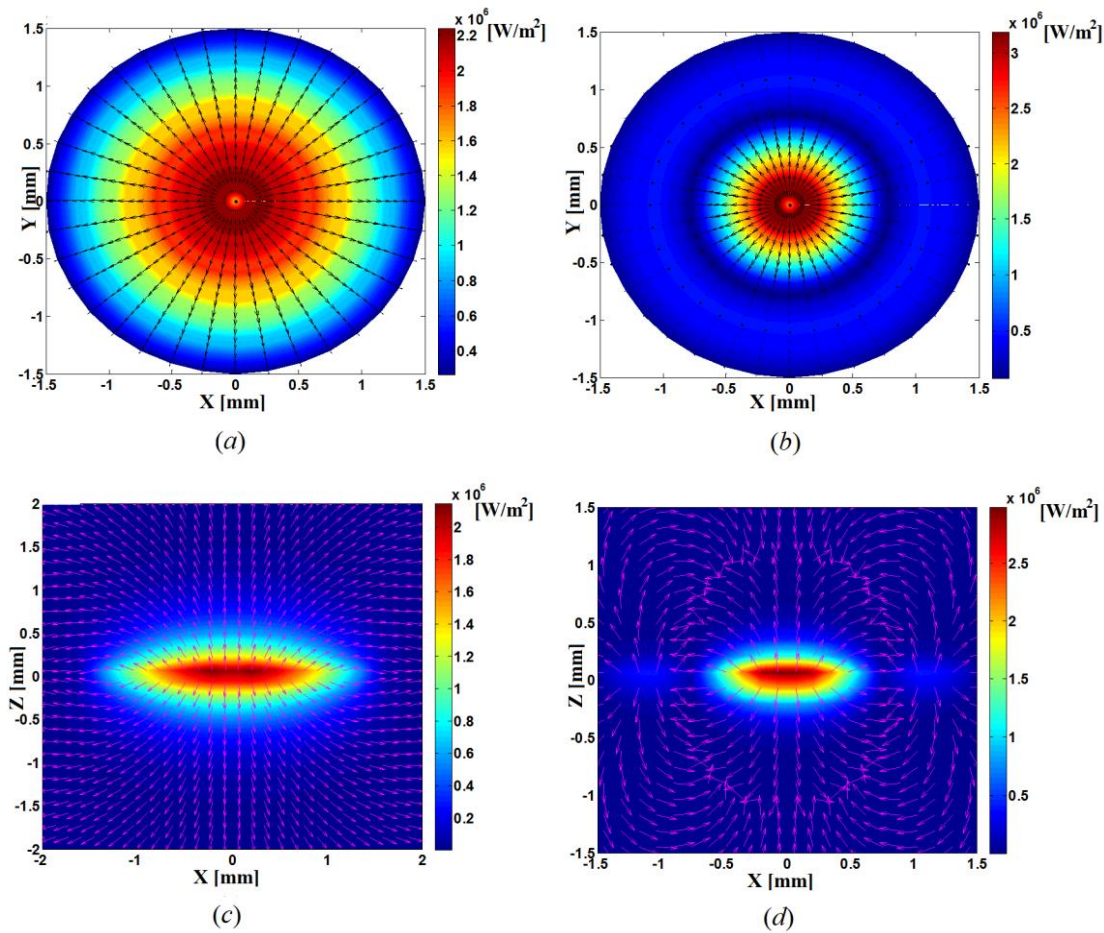

**Fig. 14. The quantity of the reactive power flow $\operatorname{Im} \vec{E}_p \times \vec{H}^*$ outside a ferrite disk. (*a*) and (*b*) The distributions on a vacuum *xy* plane 20 um above a ferrite disk for the 1st and 2nd MDMs, respectively; (*c*) and (*d*) the distributions on a cross-sectional *xz* plane for the 1st and 2nd MDMs, respectively. The arrows are unit vectors.**

In Section IV of the paper we discuss more in details the obtained analytical results on the ME-field topology.

## III. VERIFICATION OF THE ANALYTICAL RESULTS BY NUMERICAL STUDIES

Our analytical results of the ME-field structure are well verified by numerical studies based on the HFSS simulation program. Comparison between the analytical and numerical studies of the field components, shown in Figs. 15 and 16 for the 1st and 2nd MDMs, gives good correspondence between these results. In these figures, the analytically derived electric fields are potential fields calculated based on Eq. (24).

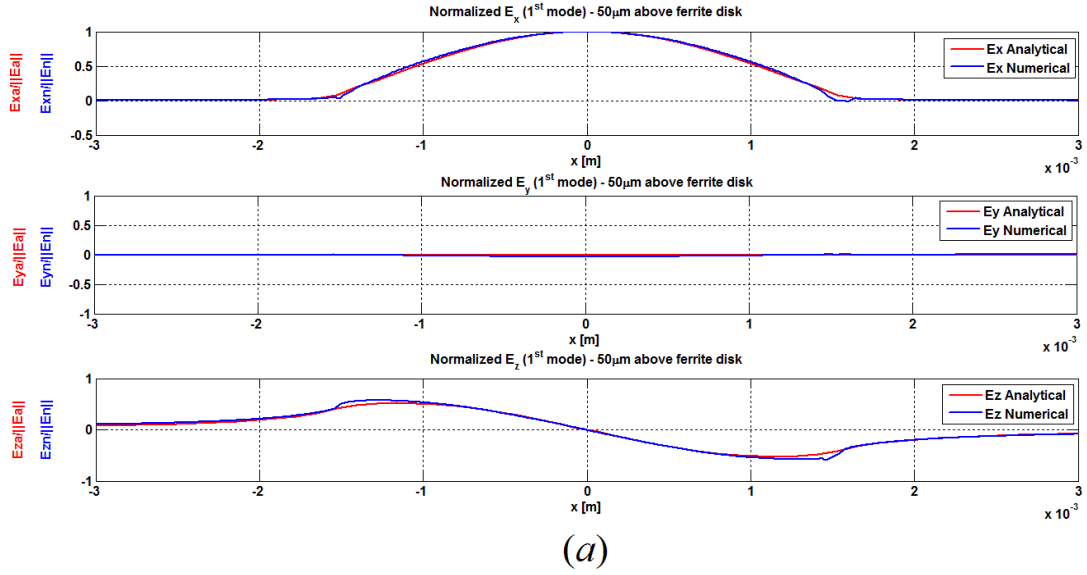

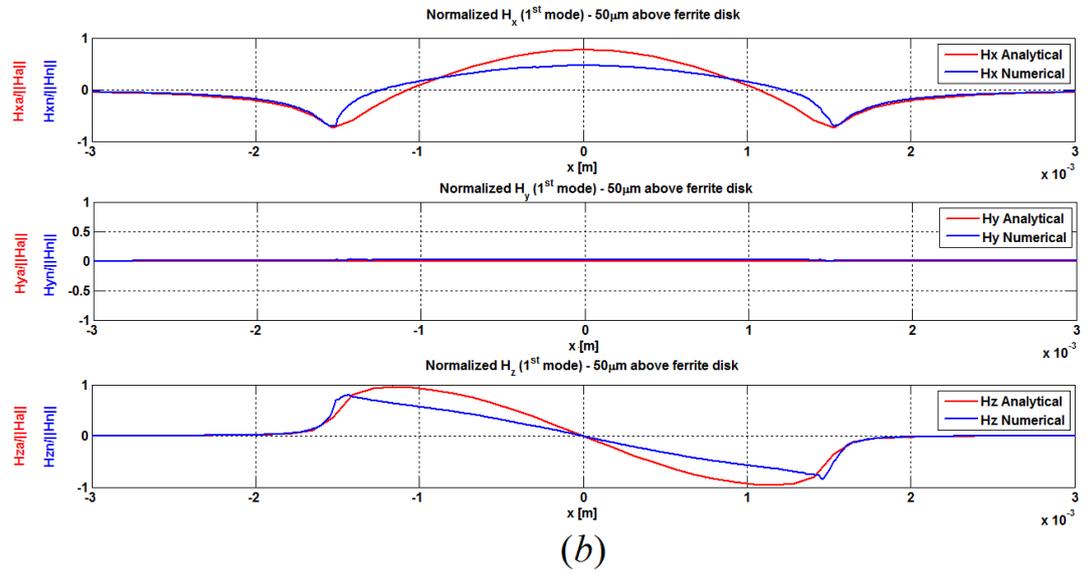

**Fig. 15.** Comparison between the analytical and numerical results of the field components for the 1st MDM. The analytically derived electric fields are potential fields calculated based on Eq. (24). The field components are shown as the quantities normalized to the modulus of the field vectors. (*a*) The electric-field components; (*b*) the magnetic-field components. A comparative analysis is made for the field components on a vacuum *xy* plane 50 um above a ferrite disk.

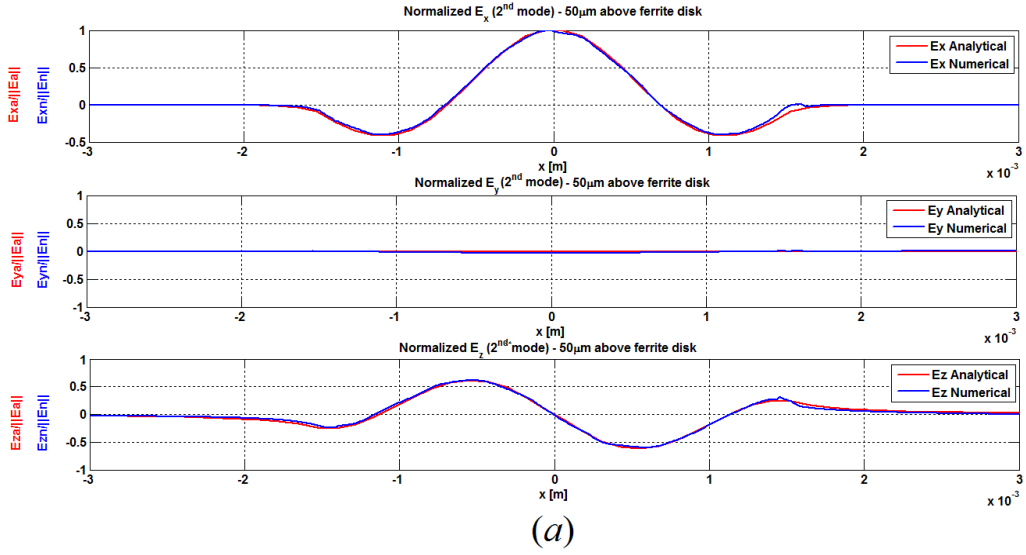

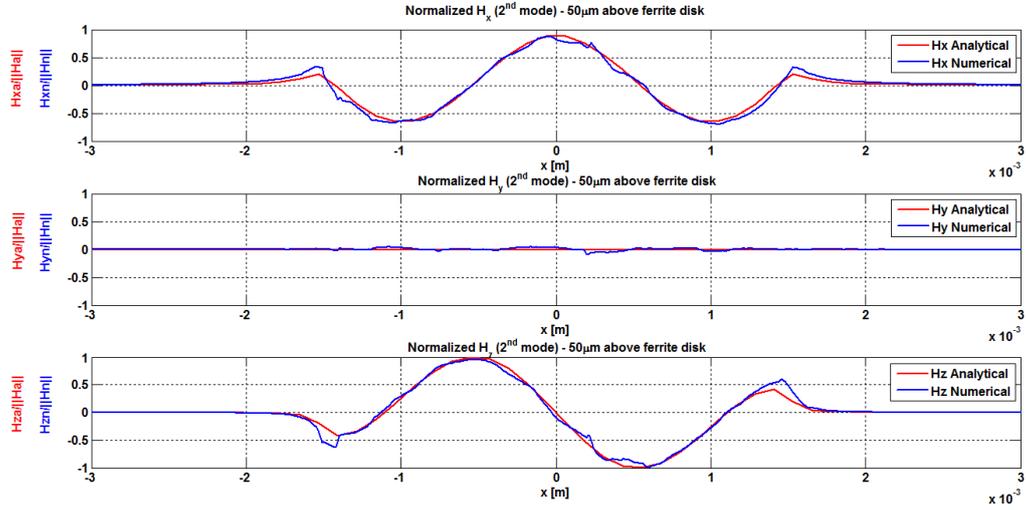

Fig. 16. Comparison between the analytical and numerical results of the field components for the 2$^{nd}$ MDM. The analytically derived electric fields are potential fields calculated based on Eq. (24). The field components are shown as the quantities normalized to the modulus of the field vectors. (*a*) The electric-field components; (*b*) the magnetic-field components. A comparative analysis is made for the field components on a vacuum *xy* plane 50 um above a ferrite disk.

The field components are shown as the quantities normalized to the modulus of the field vectors. As an illustrative example of the ME-field topology obtained based on both the theoretical and numerical analyses, we show in Fig. 17 the normalized ME-field helicity density distributions numerically calculated for the 1$^{st}$ and 2$^{nd}$ MDMs.

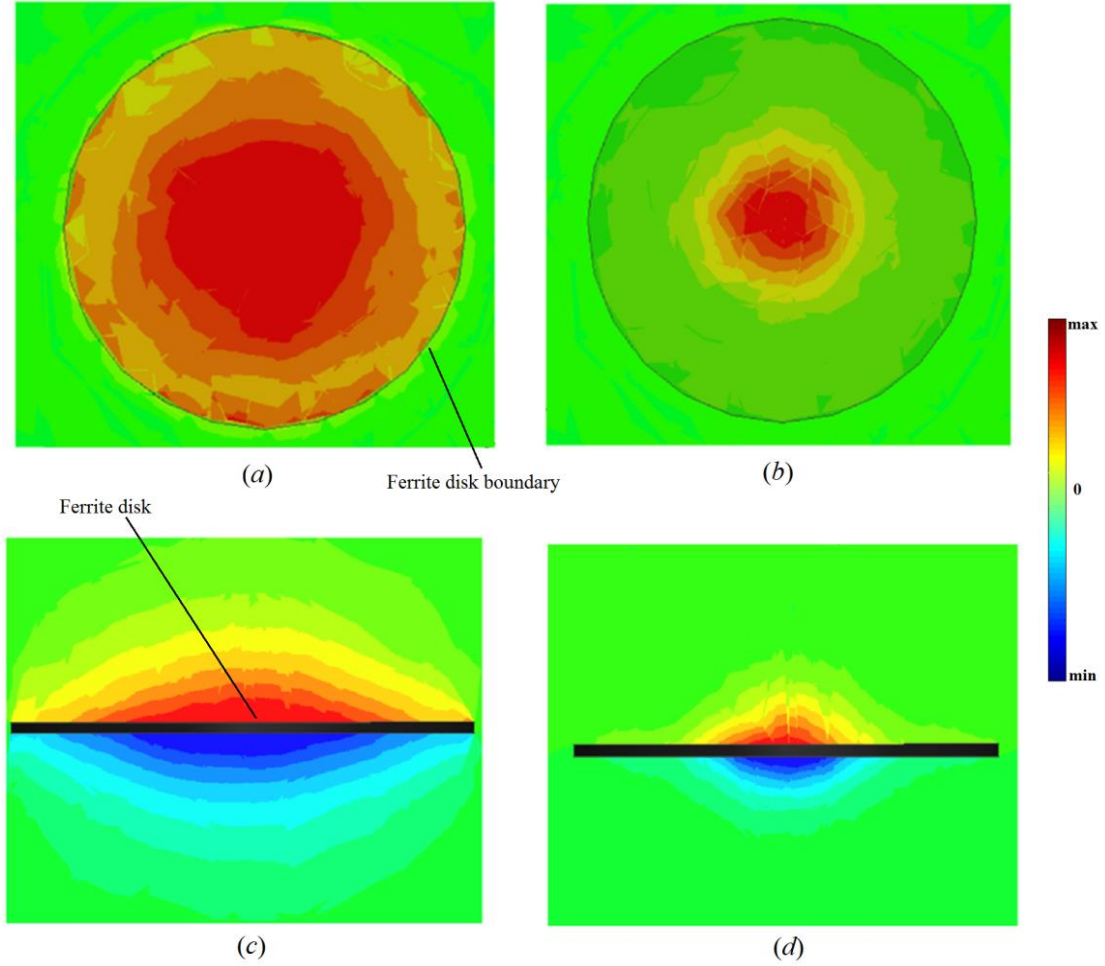

**Fig. 17. The ME-field helicity density distributions calculated numerically.** (*a*) **and** (*b*) **The distributions on a vacuum *xy* plane 20 um above a ferrite disk for the 1st and 2nd MDMs, respectively;** (*c*) **and** (*d*) **the distributions on a cross-sectional *xz* plane for the 1st and 2nd MDMs, respectively.**

One can compare these distributions with the analytically calculated ME-field helicity density distributions shown in Fig. 9. Evidently, for such a topological parameter, there is good correspondence between the two types of analyses. We restrict now further verification of the obtained analytical results by proper references of our previous numerical studies. The active-power-flow vortices shown in Figs. 11 and 13 are well verified by such vortices obtained numerically in Refs. [14, 15, 20, 21, 32]. The reactive power flows $\operatorname{Im}\vec{E}_p \times \vec{H}^*$ shown in Fig. 14 are in good correspondence with the numerical results of the reactive power flows in Ref. [21]. The analytical studies, however, display some more important features that are non-observable in the numerical results. We discuss this in the next Section of the paper.

## IV. DISCUSSION

The question of the active power flows of ME fields is not comprehensively clear. The real power flow density for MDMs expressed by Eq. (26), is obtained based on energy balance equation for monochromatic MS waves in a magnetic medium with small losses [12]. Following a formal analysis made in Ref. [33], one can conclude that such real power flow density for MDMs is related to a real part of the vector product of the curl electric field and potential magnetic field. At the same time, as we analyzed above, such an expression is not related to the electromagnetic

Poynting vector. Our analytical results reveal also some more interesting features. When one compares the active power flow distributions in Figs. 11 and 13, one finds evident similarity between these pictures. It means that there is similarity between the quantities $\text{Re}\,\vec{E}_c \times \vec{H}^*$ and $\text{Re}\,\vec{E}_p \times \vec{H}^*$. So, the roles of the curl electric field and potential electric field in the active power flow distributions are almost indistinguishable. Certainly, in the numerical studies we are dealing with the total electric fields, without any separation to the curl and potential parts.

Another interesting question concerns the reactive power flows of ME fields. As we have shown above, there are two types of reactive power flows. The first one, defined by Eq. (29), is shown in Fig. 12, while the second one, defined as $\text{Im}\,\vec{E}_p \times \vec{H}^*$, is shown in Fig. 14. One can see that these two types of the reactive power flows are localized at different parts of the ferrite disk and have different directions with respect to the disk. While in the picture shown in Fig. 12 we have the drain-type vector orientations, for the picture shown in Fig. 14 there are the source-type vector orientations. At the same time, it is worth noting that the magnitude of the reactive power flow in Fig. 12 is sufficiently smaller than such a magnitude in Fig. 14. This fact can explain why in the numerical analysis in Ref. [21] we observe only the reactive power flow defined as $\text{Im}\,\vec{E}_p \times \vec{H}^*$.

A very important issue becomes evident when one analyses the analytical results of the normalized ME-field helicity, $\cos\alpha_p$, shown in Fig. 10. In fact, there are the distributions of an angle between the potential electric and magnetic fields. From Fig. 10 (*a*) one can see that for the 1st MDM, the potential electric and magnetic fields in an entire vacuum region are mutually parallel (above a ferrite disk) or mutually anti-parallel (below a ferrite disk). For the 2nd MDM [see Fig. 10 (*b*)], above a ferrite disk there are the regions both with mutually parallel and anti-parallel electric and magnetic fields. Between these two regions, there is an intermediate area of mutual orientations of the field vectors. Similar distributions one has below a ferrite disk. Now the question arises: Whether such distributions of the potential electric and magnetic fields can be considered as the fields originated from a Tellegen particle? Tellegen considered an assembly of electric-magnetic dipole twins, all of them lined up in the same fashion (either parallel or anti-parallel) [34]. Since 1948, when Tellegen suggested such "glued pairs" as structural elements for composite materials, electromagnetic properties of these complex media was a subject of serious theoretical studies (see, e.g. Refs. [35 – 37]). Till now, however, the problem of creation of the Tellegen medium is a subject of strong discussions. The question, whether the Tellegen particles really exist in electromagnetics, is still open. The electric polarization is parity-odd and time-reversal-even. At the same time, the magnetization is parity-even and time-reversal-odd [31]. These symmetry relationships make questionable an idea that a simple combination of two (electric and magnetic) small dipoles can give the local cross-polarization (magnetoelectric) effect. In our case of the ME fields we do not have properties of the cross (magnetoelectric) polarizabilities. The ME-coupling properties are originated from magnetization dynamics of MDM oscillations in a quasi-2D ferrite disk. These oscillations are macroscopically coherent quantum states, which experience broken mirror symmetry and also broken time-reversal symmetry [12, 13]. The potential electric and magnetic field of the ME-field structure are the fields originated from the topological properties of magnetization [15].

It is also worth noting here that the normalized ME-field helicity calculated based on Eq. (25) and shown in Fig. 10, is different from such a parameter shown in the numerical results [38 – 40]. In the numerical investigations, one cannot separate the ME fields from the external EM radiation. Also, one cannot separate the potential and curl fields. It means that in the numerical studies, the normalized ME-field helicity is calculated as $\cos\alpha = \frac{\text{Re}\left(\vec{E}_p \cdot \vec{H}^*\right)}{|\vec{E}||\vec{H}|}$, where the fields $\vec{E}$ and $\vec{H}$ contain both the potential and curl components of the external-radiation EM and eigen-oscillation ME fields. The analytical and numerical results on the normalized ME-field helicity are essentially different in vacuum regions, where a magnitude of the potential electric field is reduced (the vacuum regions sufficiently far from a ferrite disk and the vacuum regions which are peripheral with respect to the disk center).

## V. CONCLUSION

Since MDM oscillations are energetically orthogonal (*G* modes), one has the same positions of the spectral peaks (with respect to a signal frequency at a constant bias magnetic field or with respect to a bias magnetic field at a constant signal frequency) in different microwave structures with an embedded quasi-2D ferrite disk. At the same time, in such different structures there are different numerically observed topological characteristics of the microwave fields. While in numerical investigations, one cannot separate the ME fields from the external EM radiation, the theoretical analysis allows clearly distinguish the eigen topological structure of ME fields. In the present paper, we obtained analytically topological characteristics of the ME-field modes. For this purpose, we used a method of successive approximations. Based on the analytical method, we have shown a "pure" structure of the electric and magnetic fields outside a MDM ferrite disk. We analyzed theoretically the fundamental topological characteristics, which are not observed in the numerical results. The analytical studies of topological properties of the ME fields can be useful for novel near- and far-field microwave applications. Strongly localized ME fields open unique perspective for sensitive microwave probing of structural characteristics of chemical and biological objects. The presence of a biological sample with chiral properties will necessarily alter the near-field topology which in turn will change the spectral characteristics of the MDM ferrite disk.

## References

[1] L. D. Landau and E. M. Lifshitz, *Electrodynamics of Continuous Media*, 2nd ed (Pergamon, Oxford, 1984).
[2] D. C. Mattis, *The Theory of Magnetism* (Harper & Row Publishers, New York, 1965).
[3] A. I. Akhiezer, V. G. Bar'yakhtar, and S. V. Peletminskii, *Spin Waves* (North-Holland, Amsterdam, 1968).
[4] A. G. Gurevich and G. A. Melkov, *Magnetic Oscillations and Waves* (CRC Press, New York, 1996).
[5] L. R. Walker, Phys. Rev. 105, 390 (1957).
[6] R. L. White and I. H. Solt, Jr., Phys. Rev. **104**, 56 (1956).
[7] J. F. Dillon, Jr., J. Appl. Phys. **31**, 1605, (1960).
[8] T. Yukawa and K. Abe, J. Appl. Phys. **45**, 3146 (1974).
[9] E. O. Kamenetskii, A. K. Saha, and I. Awai, Phys. Lett. **A 332**, 303 (2004).
[10] E. O. Kamenetskii, Phys. Rev. E **63**, 066612 (2001).
[11] E. O. Kamenetskii, M. Sigalov, and R. Shavit, J. Phys.: Condens. Matter **17**, 2211 (2005).
[12] E. O. Kamenetskii, J. Phys. A: Math. Theor. **40**, 6539 (2007).
[13] E. O. Kamenetskii, J. Phys.: Condens. Matter **22**, 486005 (2010).
[14] E. O. Kamenetskii, R. Joffe, and R. Shavit, Phys. Rev. A **84**, 023836 (2011).
[15] E. O. Kamenetskii, R. Joffe, and R. Shavit, Phys. Rev. E **87**, 023201 (2013).
[16] D. M. Lipkin, J. Math. Phys. **5**, 696 (1964).

[17] Y. Tang and A. E. Cohen, Phys. Rev. Lett. **104**, 163901 (2010).
[18] K. Bliokh and F. Nori, Phys. Rev. A **83**, 021803(R) (2011).
[19] J. S. Choi and M. Cho, Phys. Rev. A **86**, 063834 (2012).
[20] E. O. Kamenetskii, M. Sigalov, and R. Shavit, Phys. Rev. A **81**, 053823 (2010).
[21] E.O. Kamenetskii, M. Berezin, R. Shavit, arXiv:1502.00220.
[22] M. Fiebig, J. Phys D: Appl. Phys. **38**, R123 (2005).
[23] M. Mostovoy, Phys. Rev. Lett. **96**, 067601 (2006).
[24] N. A. Spaldin, M. Fiebig, and M. Mostovoy, J. Phys.: Condens. Matter **20**, 434203 (2008).
[25] T. Thonhauser, D. Ceresoli, D. Vanderbilt, and R. Resta, Phys. Rev. Lett. **95**, 137205 (2005).
[26] D. Ceresoli, T. Thonhauser, D. Vanderbilt, and R. Resta, Phys Rev. B **74**, 024408 (2006).
[27] A. M. Essin, J. E. Moore, and D. Vanderbilt, Phys. Rev. Lett. **102**, 146805 (2009).
[28] A. Malashevich, I. Souza, S. Coh, and D. Vanderbilt, New J. Phys. **12**, 053032 (2010).
[29] S. Coh, D. Vanderbilt, A. Malashevich, and I. Souza, Phys. Rev. B **83**, 085108 (2011).
[30] F. Wilczek, Phys. Rev. Lett. **58**, 1799 (1987).
[31] J. D. Jackson, *Classical Electrodynamics*, 2nd ed (Wiley, New York, 1975).
[32] M. Sigalov, E. O. Kamenetskii, and R. Shavit, J. Phys.: Condens. Matter **21**, 016003 (2009).
[33] D. D. Stancil, *Theory of Magnetostatic Waves* (Springer-Verlag, New York, 1992).
[34] B. D. H. Tellegen, Philips. Res. Rep. **3**, 81 (1948).
[35] E. J. Post, *Formal Structure of Electromagnetics* (Amsterdam, North-Holland, 1962).
[36] I. V. Lindell, A. H. Sihvola, S. A. Tretyakov and A J Viitanen, *Electromagnetic Waves in Chiral and Bi-Isotropic Media* (Boston, MA, Artech House, 1994).
[37] A. Lakhtakia, *Beltrami Fields in Chiral Media* (Singapore, World Scientific, 1994).
[38] M. Berezin, E. O. Kamenetskii, and R. Shavit, J. Opt. **14**, 125602 (2012).
[39] R. Joffe, E. O. Kamenetskii and R. Shavit, J. Appl. Phys. **113**, 063912 (2013).
[40] M. Berezin, E. O. Kamenetskii, and R. Shavit, Phys. Rev. E 89, 023207 (2014).